\documentclass[pra,twocolumn,showpacs]{revtex4}
\usepackage{amsmath}
\usepackage{graphicx}
\usepackage{amssymb} 

\begin{document} 
\bibliographystyle{prsty} 
 
\title{The Groverian Measure of Entanglement for Mixed States}
 
\author{Daniel Shapira, Yishai Shimoni and Ofer Biham} 
\affiliation{Racah Institute of 
Physics, The Hebrew University, Jerusalem 91904, Israel} 
 
\begin{abstract} 
The Groverian entanglement measure introduced earlier
for pure quantum states 
[O. Biham, M.A. Nielsen and T. Osborne, Phys. Rev. A 65, 062312 (2002)]
is generalized to the case of mixed states, in a way that maintains
its operational interpretation. 
The Groverian measure of a mixed state of $n$ qubits is obtained by a
purification procedure into a pure state of $2n$ qubits, followed by
an optimization process 
before the resulting state is fed into Grover's search algorithm.
The Groverian measure, expressed in terms of the maximal 
success probability of the algorithm, 
provides an operational measure of entanglement
of both pure and mixed quantum states of multiple qubits.
These results may provide further insight into the role of entanglement in
making quantum algorithms powerful. 
\end{abstract} 
 
 
\maketitle 
 
\section{Introduction} 
\label{sec:introduction}

The potential speedup offered by quantum computers is exemplified
by Shor's factoring algorithm
\cite{Shor1994}, Grover's search algorithm 
\cite{Grover1996,Grover1997a},
and algorithms for
quantum simulation
\cite{Nielsen2000}.
Although the origin of 
this speed-up is not fully understood, 
there are indications that quantum entanglement 
plays a crucial role
in making quantum algorithms efficient
\cite{Jozsa2003,Vidal2003}. 
In particular,
it was shown that quantum algorithms that do not create entanglement 
can be simulated efficiently on a classical computer
\cite{Aharonov1996}.
It is therefore of interest to quantify the entanglement 
produced by quantum algorithms and examine its correlation
with their efficiency.  
This requires to develop entanglement measures for the quantum states
of multiple qubits that appear in quantum algorithms.
These include pure states as well as mixed states, which would 
inevitably show up when decoherence effects are taken into account.

The special case of bi-partite entanglement
has been studied extensively in recent years and suitable
entanglement measures were introduced.
It was established that bi-partite entanglement can be considered
as a resource for teleportation
\cite{Bennett1993}. 
The entanglement of bi-partite pure states can be evaluated by the
von Neumann entropy of the reduced density matrix,
traced over one of the parties.
For bi-partite mixed states, several measures were proposed
\cite{Bennett1996a,Bennett1996b,Popescu1997} and for the special case
of states of two qubits an exact formula for the
entanglement of formation was obtained
\cite{Hill1997,Wootters1998}.
For mixed states of multiple qubits, 
entanglement measures based on distance measures in Hilbert 
space were proposed
\cite{Vedral1997,Vedral1997a,Vedral1998}.

Consider a mixed quantum state $\rho$ of $n$ qubits.
The state is non-entangled, or separable, 
if its density matrix can be written in the form

\begin{equation}
\rho = \sum_\mu P_\mu\rho^1_\mu\otimes\dots\otimes\rho^n_\mu,
\label{eq:separable}
\end{equation}

\noindent
where 
$\rho^k_\mu$, $k=1,\dots,n$ 
is a density operator of a pure state of the $k$th qubits,
namely
$\rho^k_\mu = | \psi^k_\mu \rangle \langle \psi^k_\mu |$ 
and 
$\sum_\mu P_\mu=1$.
In the special case  
that $\rho$ is
a pure state, 
all probabilities vanish except for
$P_1=1$, and the state
can be expressed by

\begin{equation}
| \psi \rangle=|\psi^1\rangle\otimes\dots\otimes|\psi^n\rangle.
\label{eq:pureNE}
\end{equation}

\noindent
Such states are called tensor-product states.
In order to evaluate the entanglement of a quantum state, $\rho$,
one needs a scalar function 
$E(\rho)$ 
[or $E(\psi)$ for pure states] 
called an {\em entanglement measure}
that satisfies 
\cite{Vedral1997,Vedral1997a,Vedral1998,Vidal2000,Horodecki2000}:
(a) $E(\rho)=0$ if and only if $\rho$ is a separable state;
(b) Assuming that each qubit is held by a different party,
it is not possible to increase $E(\rho)$ by local operations and classical
communication (LOCC) between the parties.
Consider the special case of local unitary operators.
Such operators cannot decrease
$E(\rho)$ because if they could 
then the inverse operators (which are also unitary) would increase it
and thus contradict the second condition above.
The conclusion is that local unitary operators cannot change 
$E(\rho)$.

The Groverian entanglement measure, $G(\psi)$, 
provides an operational measure of entanglement
for pure states of multiple qubits
\cite{Biham2002}. 
It is related to the success probability of Grover's search
algotirhm when the state 
$| \psi \rangle$
is used as the initial state. 
A pre-precessing stage is allowed 
in which an arbitrary local unitary operator is applied to each
qubit.
These operators are optimized in order to
obtain the maximal success probability 
of the algorithm,
$P_{\rm max}(\psi)$.
The Groverian measure is given by 
$G(\psi) = \sqrt{1 - P_{\rm max}(\psi)}$. 
The Groverian measure was used in order to evaluate
the entanglement in certain quantum states of high symmetry
as well as in states that are generated during the operation
of quantum algorithms
\cite{Shimoni2004}.
For example, it was found that Grover's iterations generate highly
entangled intermediate states, even in case that the initial and
the final states are product states.

In this paper we generalize the Groverian entanglement measure to
the case of mixed states.
The Groverian measure, 
$G(\rho)$,
of a given mixed state
$\rho$, of $n$ qubits, is obtained by its purification into a pure state
of $2n$ qubits. 
An optimization procedure based on Uhlmann's theorem
\cite{Uhlmann1976}
is then applied before the resulting pure state is fed into Grover's
algorithm.
$G(\rho)$ is then expressed in terms of
the maximal success probability 
$P_{\rm max}(\rho)$,
as described above for pure states.

In Sec. \ref{sec:algorithm} we briefly describe Grover's search algorithm,
in a context suitable to this paper. 
In Sec. \ref{sec:pure} we review the Groverian entanglement measure
for pure states.
The generalization to mixed states is presented in 
Sec. \ref{sec:mixed} 
and its operational interpretation is considered. 
The results are summarized and discussed in 
Sec.
\ref{sec:summary}.

\section{Grover's Search Algorithm}
\label{sec:algorithm}

Consider a search space $D$ containing $N$ elements.  We assume, for 
convenience, that $N = 2^n$, where $n$ is an integer. This way, 
the elements of $D$ can be represented by an $n$-qubit register
$| x \rangle = | x_1,x_2,\dots,x_n \rangle$, 
with the computational basis states 
$| i \rangle$, $i=0,\dots,N-1$.  
We assume that one 
element in the search space is marked, namely it is the
solution of the search problem.  
The distinction between the marked 
and unmarked elements is expressed by a suitable function, 
$f: D \rightarrow \{0,1\}$, 
such that $f=1$ for the marked element, and $f=0$ for the rest. 
The search for the marked element now becomes a search for the element 
for which $f=1$.  
To solve this 
problem on a classical computer one needs to evaluate $f$ for each 
element, one by one, until a marked state is found.  Thus, on average, 
$N/2$ evaluations of $f$ are required and $N$ in the worst case.
For a quantum computer, on which
$f$ is evaluated 
\emph{coherently}, 
it was shown that a sequence of unitary operations,
called Grover's algorithm
and denoted by
$U_G$, 
can locate the marked element using only $O(\sqrt{N})$ coherent 
queries of $f$ \cite{Grover1996,Grover1997a}.  

Starting with the equal superposition state, 

\begin{equation}
|\eta\rangle = \frac{1}{\sqrt{N}} \sum_{i=0}^{N-1}|i\rangle,
\end{equation}

\noindent 
and applying the operator $U_G$
one obtains
$U_G |\eta\rangle = |m\rangle + O({1}/{N})$,
where 
$| m \rangle$
is the marked state.
Thus,
the success probability of the algorithm is almost unity
\cite{Grover1996,Grover1997a}.
With this performance, 
Grover's algorithm was shown to be optimal
\cite{Zalka1999}
namely, it is as efficient as theoretically possible
\cite{Bennett1997}. 
The adjoint equation takes the form

\begin{equation}
\langle\eta| = \langle m|U_G + O({1}/{N}),
\label{eq:<mU}
\end{equation}

\noindent
where the error is due to the discreteness of the Grover iterations
\cite{Shapira2005}.
If an arbitrary pure state, $|\psi\rangle$,
is used as the initial state 
instead of the state $| \eta \rangle$,
the success probability is reduced
\cite{Biron1998,Biham1999}.
It is given by
\cite{Biham2003}

\begin{equation}
P_s(\psi) = 
|\langle m|U_G|\psi\rangle|^2 + O({1}/{N}).
\label{eq:Psmpsi}
\end{equation}

\noindent
Using Eq.~(\ref{eq:<mU}) we obtain

\begin{equation}
P_s(\psi) = |\langle\eta|\psi\rangle|^2 + O({1}/{N}),
\label{eq:etaU}
\end{equation}

\noindent
namely, the success probability is determined by the overlap
between 
$| \psi \rangle$
and 
$| \eta \rangle$.

\section{The Groverian Measure for Pure States}
\label{sec:pure}

\begin{figure}
\includegraphics[angle=270, width=8cm]{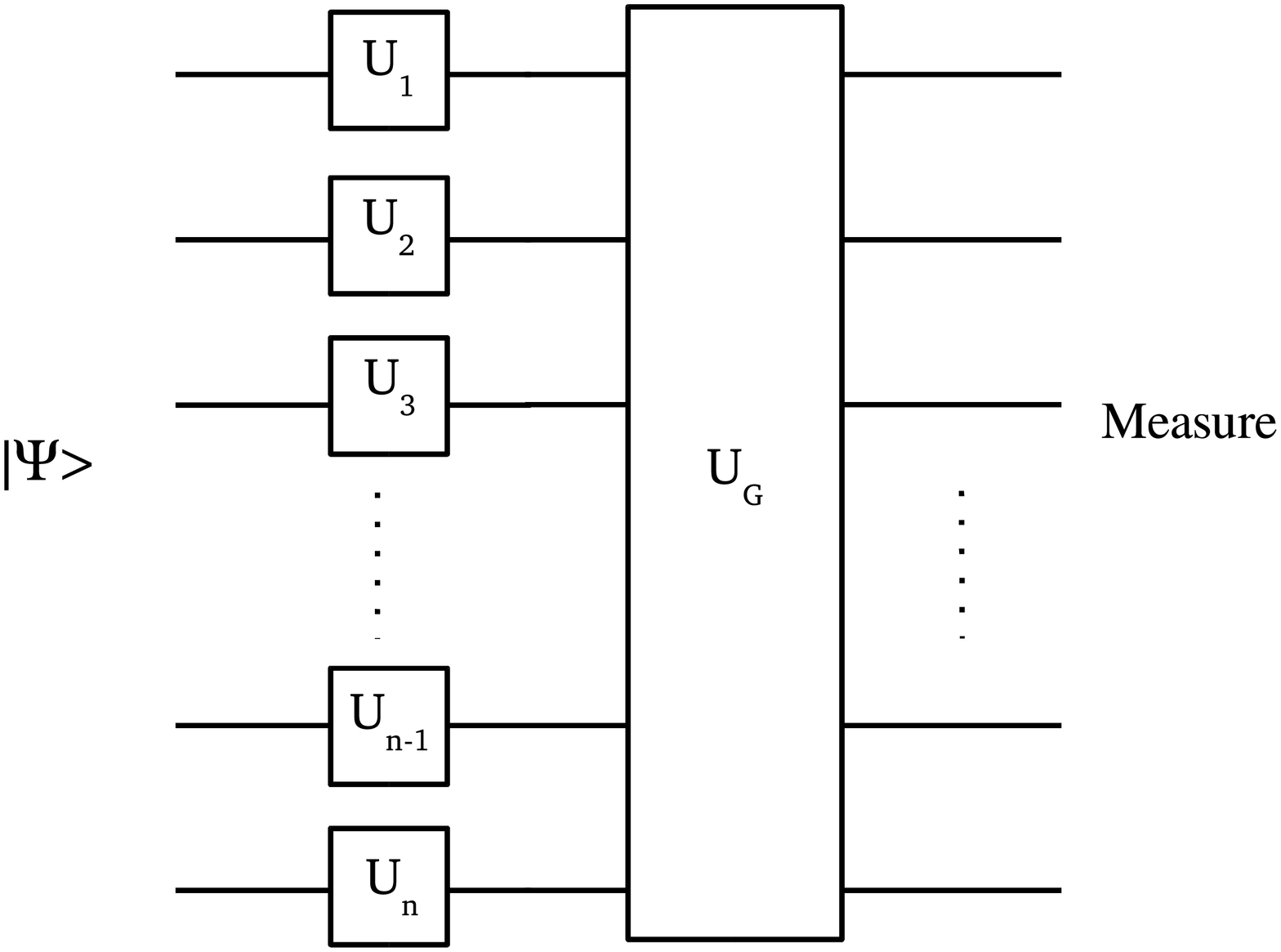}
\caption{
The quantum circuit that exemplifies the operational 
meaning of $G(\psi)$.
An arbitrary pure state 
$| \psi \rangle$
of $n$ qubits is inserted as the input state.
In the pre-processing stage, a local unitary operator
is applied to each qubit before the resulting state
is fed into Grover's algorithm.
The local unitary operators
$U_i$, $i=1,\dots,n$
are optimized in order to maximize the success 
probability of the search algorithm for the given
initial state
$| \psi \rangle$.
}
\label{fig1}
\end{figure}

Consider Grover's search algorithm, in which an arbitrary pure state
$| \psi \rangle$ is used as the initial state.
Before applying the operator $U_G$, there is a pre-processing stage
in which arbitrary local unitary operators,
$U_1$, $U_2$, $\dots$, $U_n$,
are applied on the $n$ qubits in the register
(Fig. \ref{fig1}).
These operators are chosen such that the success probability
of the algorithm would be maximized.
The maximal success probability is thus given by

\begin{equation}
P_{\rm max}(\psi) = 
\max_{U_1,U_2,\dots,U_n}
|\langle m|U_G(U_1\otimes\dots\otimes U_n)|\psi\rangle|^2.
\label{eq:Pmax}
\end{equation}

\noindent
Using Eq.~(\ref{eq:<mU}), this can be re-written as

\begin{equation}
P_{\rm max}(\psi) = 
\max_{U_1,U_2,\dots,U_n}|\langle\eta|U_1\otimes\dots\otimes U_n|\psi\rangle|^2,
\end{equation}

\noindent
or
$P_{\rm max}(\psi) = 
\max_{|\phi\rangle \in T}|\langle\phi|\psi\rangle|^2$,
where $T$ is the space of all tensor product states of the form
$|\phi\rangle = |\phi_1\rangle\otimes\dots\otimes|\phi_n\rangle$.
The Groverian measure is given by
\cite{Biham2002}

\begin{equation}
G(\psi) = 
\sqrt{1-\max_{|\phi\rangle \in T}|\langle\phi|\psi\rangle|^2}.
\label{eq:G(psi)}
\end{equation}

\noindent
For the case of pure states, for which $G(\psi)$
is defined, it is closely related to an entanglement measure 
introduced in Refs.
\cite{Vedral1997,Vedral1997a,Vedral1998} and
was shown to be an entanglement monotone.
The latter measure is defined for both pure and mixed 
states. 
It can be interpreted as the distance 
between the given state and the nearest separable state
and expressed in terms of the fidelity of the
two states.
Based on these results, it was shown 
\cite{Biham2002}
that $G(\psi)$
satisfies:
(a) $G(\psi) \geq 0$, with equality only when $|\psi\rangle$
is a product state;
(b) $G(\psi)$ cannot be increased using local operations 
and classical communication
(LOCC). 
Therefore, $G(\psi)$ is an entanglement monotone
for pure states.
A related result was obtained in Ref.
\cite{Miyake2001},
where it was shown that the evolution of the quantum state  
during the iteration of Grover's
algorithm corresponds
to the shortest path in Hilbert space
using a suitable metric.

\section{The Groverian Measure for mixed states}
\label{sec:mixed}

\subsection{Desired Properties and construction of $G(\rho)$}

\begin{figure}
\includegraphics[width=8cm]{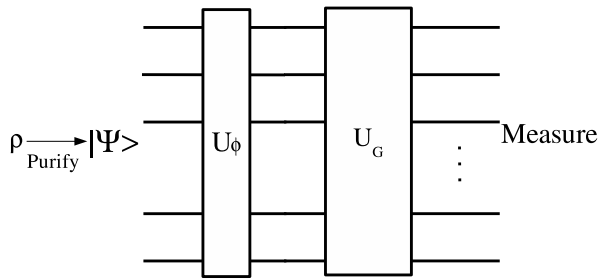}
\caption{
The quantum circuit for the Groverian measure $G(\rho)$ of a mixed state
of $n$ qubits.
The given state $\rho$ 
is purified into a state $| \psi \rangle$
of $2n$ qubits.
In the pre-processing stage, a unitary operator
$U_\phi$ 
is applied on $| \psi \rangle$,
before the resulting state is fed into Grover's search algorithm
in the space of $2n$ qubits.
The operator
$U_\phi$ 
belongs to a class of operators that their adjoint operators
satisfy
$U_\phi^{\dagger} | \eta \rangle = | \phi \rangle$
where 
$| \phi \rangle$ 
is a purification of a separable state of $n$ qubits.
The choice of
$U_\phi$ 
is optimized
in order to maximize the success probability of the
search process for the given state $\rho$.
}
\label{fig2}
\end{figure}

The quantum circuit that demonstrates the evaluation of the
Groverian measure $G(\rho)$ for a mixed state $\rho$ is
shown in 
Fig. \ref{fig2}.
The given state $\rho$ is purified into a pure state
$|\psi\rangle$
of $2n$ qubits. 
In the pre-processing 
an operator 
$U_{\phi}$ 
is applied.
This operator has the property that its adjoint operator satisfies
$U_{\phi}^{\dagger} | \eta \rangle = | \phi \rangle$,
where 
$| \phi \rangle$
is a purification of a separable state of $n$ qubits.
The resulting state is then inserted into Grover's search
algorithm.
An optimization is performed over the entire family
of operators 
$U_{\phi}$ 
that satisfy the above condition, in order to
maximize the success probablity of the search algorithm.
Equivalently, the optimization can be performed over all
the separable states, $\sigma$, and over all possible purifications
$| \phi \rangle$ of each one of them.
The maximal success probability for the initial state
$\rho$ is denoted by $P_{\rm max}(\rho)$
and the Groverian measure is given by
$G(\rho)=\sqrt{1-P_{\rm max}(\rho)}$.

The construction of the Groverian measure for mixed states
is based on a purification process, where
the pure state $|\psi\rangle$ of $2n$ qubits 
is a purification of the mixed state $\rho$ of $n$ qubits.
Similarly,
$|\phi\rangle$ 
is a purification of a 
separable mixed state $\sigma$.
As in the case of pure states,
we introduce a pre-processing stage before the state is inserted into
Grover's algorithm.
During pre-processing an optimization is performed over 
a certain class of unitary 
operators $U_{\phi}$.
These operators satisfy
$|\phi\rangle = U_{\phi}^{\dagger}|\eta\rangle$, 
where 
$|\phi\rangle$ 
is a purification of a separable state 
$\sigma$
of $n$ qubits.
The maximal success probability is given by

\begin{equation}
P_{\rm max}(\rho) = \max_{\sigma\in S} 
\max_{|\phi\rangle} \max_{|\psi\rangle}
\left|\langle m | U_G U_{\phi} | \psi \rangle \right|^2.
\end{equation}

\noindent
Using Eq.
(\ref{eq:<mU}) we obtain

\begin{equation}
P_{\rm max}(\rho) = \max_{\sigma\in S} 
\max_{|\phi\rangle} \max_{|\psi\rangle}
\left|\langle\eta|U_{\phi}|\psi\rangle\right|^2,
\end{equation}

\noindent
where $S$ is the set of separable states of $n$ qubits.
The maximization is over all separable states
$\sigma$ of $n$ qubits, and for each of them, over all possible
purifications $| \phi \rangle$ of $2n$ qubits.
This can be rewritten as

\begin{equation}
P_{\rm max}(\rho) = 
\max_{|\phi\rangle} \max_{|\psi\rangle}
\left|\langle\phi|\psi\rangle\right|^2.
\end{equation}

\noindent
The first
maximization is over all possible 
states 
$|\phi\rangle$ of $2n$ qubits 
which are
purifications of separable states $\sigma$ of $n$ qubits.
The second maximization is over all states 
$| \psi \rangle$
of $2n$ qubits which are purifications of $\rho$.
According to Uhlmann's theorem,
the fidelity of any two states $\rho$ and $\sigma$ 
of $n$ qubits satisfies

\begin{equation}
F(\rho,\sigma) = \max_{|\phi\rangle}\max_{|\psi\rangle}
|\langle\phi|\psi\rangle|^2,
\end{equation}

\noindent
where 
$|\phi\rangle$ 
and 
$|\psi\rangle$,
of $2n$ qubits, 
are purifications of 
$\rho$ and $\sigma$ respectively
\cite{Uhlmann1976}. 
A useful 
corollary 
(presented in Excercise 9.15 on page 411 of Ref.  
\cite{Nielsen2000})
enables us to remove the optimization on $|\psi \rangle$,
leading to

\begin{equation}
F(\rho,\sigma)=\max_{|\phi\rangle}|\langle\phi|\psi\rangle|^2,
\end{equation}

\noindent
where $|\psi\rangle$ is an arbitrary purification of $\rho$.
Using this corollary we find that

\begin{equation}
G(\rho) = \sqrt{1-\max_{\sigma\in S}F(\rho,\sigma)}.
\label{eq:Grho}
\end{equation}

\noindent
An entanglement measure for mixed states, 
$\rho$,
should satisfy the conditions:
(a)
$G(\rho) \geq 0$, where the equality is obtained only if $\rho$
is a separable state;
(b)
$G(\rho)$ cannot be increased using LOCC. 
To express the second condition 
we introduce a 
complete set of operators, defined by 
$\{M_i\}_{i=1}^m$
where 
$M_i=M_i(1)\otimes M_i(2)\otimes\dots\otimes M_i(n)$
and
$\sum_{i=1}^m M_iM_i^{\dagger}=I$.
In this notation,
the condition is that for every density matrix $\rho$

\begin{equation}
G\left( \sum_{i=1}^m M_i\rho M_i^{\dagger} \right) \leq G(\rho).
\end{equation}

\noindent
We proceed to prove that $G(\rho)$ fulfills the two conditions 
described above:
As shown in Refs.
\cite{Vedral1997,Vedral1997a,Vedral1998},
functions of the form of $G(\psi)$ satisfy the
conditions above for an entanglement monotone.
More specifically, 
there exist two specific purifications, 
$|\phi_0\rangle$ 
and 
$|\psi_0\rangle$, 
for which 
$F(\rho,\sigma)=|\langle\phi_0|\psi_0\rangle|^2$.
Thus,
$G(\rho)=0$ if and only if 
$F(\rho,\sigma)=1$. 
In this case
$|\langle\phi_0|\psi_0\rangle|^2=1$, 
or 
$|\phi_0\rangle=e^{i\alpha}|\psi_0\rangle$,
thus $\rho=\sigma$. 
Since $\sigma$ is separable then so is $\rho$,
and the first condition is satisfied.

In order to prove the second condition 
one can use the monotonous quality of the fidelity
under trace preserving operations 
\cite{Nielsen2000}. 
This means that for every complete set of operators 

\begin{equation}
F(\rho,\sigma)\leq 
F\left(\sum_i M_i\rho M_i^\dagger,\sum_i M_i\sigma M_i^\dagger \right),
\end{equation}

\noindent
where
the separable state 
$\sigma$ 
remains separable under the transformation.
As a result,

\begin{equation}
\max_{\sigma\in S} F(\rho,\sigma)\leq 
	\max_{\sigma\in S} F\left(\sum_i M_i\rho M_i^\dagger,\sigma \right).
\end{equation}

\noindent
Therefore,
$G(\rho)$ cannot increase under such transformations. 

\subsection{The Unitary Operator $U_{\phi}$ and its Operational Interpretation}

Consider a mixed state $\sigma$ of $n$ qubits.
The state $\sigma$ can be purified 
into a pure state of $2n$ qubits, 
half of them associated with the original subspace
$Q$ and the rest with the added subspace $R$
\cite{Nielsen2000}. 
Such purification takes the form

\begin{equation}
| \phi \rangle = (U_R\otimes\sqrt{\sigma}U_Q) \sum_i|i_R\rangle|i_Q\rangle,
\end{equation}

\noindent
where 
$\{|i_R\rangle\}$ 
and 
$\{|i_Q\rangle\}$ 
are orthonormal  
basis states 
and
$U_R$ and $U_Q$ 
are unitary operators, 
in $R$ and $Q$,
respectively. 

For any 
separable mixed state,
$\sigma$ of $n$ qubits,
the unitary operators 
$U_R$ and $U_Q$
and the operator
$\sqrt{\sigma}$
provide a convenient parametrization of the 
pure state 
$|\phi\rangle$.
This follows from the fact that
Eq.~(\ref{eq:separable}) 
can also be expressed as
$\sigma=\sum_\mu \sigma^1_\mu\otimes\dots\otimes\sigma^n_\mu$,
where $\sigma^k_\mu$, $k=1,\dots,n$ 
is the density operator of a mixed state of the
$k$th qubit.
To obtain the operator 
$\sqrt{\sigma}$, one needs to diagonalize  
$\sigma$, 
according to 
$V \sigma V^{\dagger} = D$,
where 
$V$ is a suitable unitary operator,
resulting in a diagonal matrix $D$. 
Taking its square root one obtains another
diagonal matrix $d$
with matrix elements
$d_{ii}=\sqrt{D_{ii}}$. 
The process is completed by
$\sqrt{\sigma}=V^{\dagger} d V$.
The operator
$\sqrt{\sigma}$ is not a required to be unitary 
and is not used in the quantum circuit.

In order to complete the construction of the quantum circuit,
one needs the operator
$U_{\phi}^{\dagger}$
that transforms 
$| \eta \rangle$
into a purification 
$| \phi \rangle$
of a separable state
$\sigma$.
To this end we construct two bases of the space spanned
by $2n$ qubits.
The first basis is obtained from the computational basis,
$|i\rangle$ 
by appling the Hadamard transform on all qubits,
namely
$| \eta_i \rangle = H^{\otimes 2n} | i \rangle$. 
Note that 
$|\eta_0 \rangle = | \eta \rangle$.
The second basis, 
$| \phi_i \rangle$, 
can be constructed using 
the Gram-Schmidt algorithm,
starting with
$| \phi_0 \rangle = | \phi \rangle$.  
The unitary operator 
$U_{\phi}^{\dagger} = \sum_i |\phi_i\rangle\langle\eta_i|$
transforms the state
$| \eta \rangle$
into the state
$| \phi \rangle$,
which is a purification of a separable state, 
$\sigma$,
of $n$ qubits.

\section{Summary and Discussion}
\label{sec:summary}

The Groverian entanglement measure for multiple qubits, 
previously introduced for pure states, was generalized
to the case of mixed quantum states.
This generalization provides an operational measure
of entanglement for both pure and mixed states.
The operational interpretation is based on the fact that
the Groverian measure of a state $\rho$ is related to 
the success probability of Grover's search algorithm 
when $\rho$ is used as the initial state, following a 
suitable pre-processing stage.
When $\rho$ is a mixed state, the pre-processing includes
a purification procedure followed by the application of 
a certain unitary operator. 
In case that the given state $\rho$ is a pure state,
$G(\rho)$
coincides with the Groverian measure for pure states,
introduced in Ref.
\cite{Biham2002}.

It might seem surprising that in order to evaluate the
entanglement of an $n$-qubit system one needs a search space
of $2n$ qubits. 
This can be explained by considering
mixed states as open systems, 
where, in some cases, the mixture 
represents an effective entanglement with external qubits. 
Purification of the mixed state $\rho$ to $2n$ 
qubits enables us to create a closed system with no entanglement
to any external qubits, in which all the relevant information is
maintained.  
The Groverian measure for mixed states 
enables to fully capture the entanglement 
in $\rho$. 
Mathematically, this measure satisfies the 
conditions required from an entanglement 
measure. In particular, it vanishes for any separable mixed state of $n$ 
qubits, it is invariant under local unitaries and monotonically decreasing
(in the weak sense) under LOCC.

Quantum algorithms are designed to start with a well defined
initial state. The final state, just before the measurement is
taken, can be either a basis state or a superposition of 
basis states.
For example, in Grover's algorithm with a single marked state,
the desired final state (namely the marked state) is a basis state.
In Grover's algorithm with several marked states, as well as in
Shor's factoring algorithm, the desired final state is a superposition.
The analysis presented in this paper can be generalized by 
replacing Grover's algorithm by some other quantum algorithm.
If in the replacement algorithm the desired final state is 
a basis state, $G(\rho)$ will not depend on the specific algorithm.
However, for algorithms such as Shor's algorithm,
the maximal success probability may not coincide with the
expression used in the Groverian measure.
Yet, in the special case of Grover's algorithm with several
marked states, the Groverian measure still holds
\cite{Biham2003}.

Recently, 
the Groverian measure was applied in order to evaluate 
the entanglement in certain pure quantum states of multiple qubits
\cite{Shimoni2004}.
A convenient parametrization was developed that
enables analytical calculations of $G(\psi)$
for some pure states of high symmetry.
In order to evaluate it for
arbitrary states, 
a numerical minimization procedure, 
based on the steepest descent algorithm was developed.
Using this procedure, the
entanglement of intermediate states, 
generated during the evolution of Grover's algorithm,
was calculated.
It was found that even if the initial state and the target
state are product states, in intermediate stages of 
the algorithm, highly entangled states are generated,
in agreement with earlier studies in which other measures
were used
\cite{Miyake2001,Meyer2002}.
This result is interesting in the context of attempts to examine
the role of entanglement in quantum algorithms and specifically in
Grover's search algorithm. 
In particular, recent studies have shown that an implementation
of Grover's algorithm using classical media, namely, in which 
quantum entanglement does not play a role, would require an
exponentially larger overhead compared to the quantum case 
\cite{Lloyd1999,Meyer2000}.

Unlike the case of pure states of two qubits, in which
the von Neumann entropy provides a complete characterization
of the entanglement, multiple qubit states support a large 
number of different measures
\cite{Miyake2001,Meyer2002,Barnum2001,Leifer2003,Wei2003}. 
It seems that the issue of what measure is relevant 
depends on the specific physical or operational context
in which it is used.
In particular, the Groverian entanglement measure is motivated
by a quantum algorithm.
It thus appears to be a suitable measure for the evaluation of
the entanglement that is produced during the evolution of 
quantum algorithms.
The actual evaluation of entanglement measures 
turns out to be a difficult computational problem.
This is due to the fact that these measures are typically
defined as an extremum of some multi-variable function.
A singular result in this context is the explicit
formula for the entanglement of formation of mixed states
of two qubits, obtained in Refs.
\cite{Hill1997,Wootters1998}.

A related operational interpretation of the fidelity,
which is also based on Uhlmann's theorem, was introduced
in Ref.
\cite{Dodd2002}.
In that case the fidelity $F(\rho,\sigma)$
of the output states of a noisy channel provides an upper bound
on the overlap of the input states, under the assumption
that they were pure.

The generalization of the Groverian measure to mixed states
may provide further insight into the role of entanglement in
making quantum algorithms powerful.
It would be interesting to use this measure to evaluate the entanglement
generated by quantum algorithms using mixed states,
particularly when decoherence effects are taken into account.



\end{document}